\input harvmac

\font\tenmsb=msbm10

\newfam\msbfam
\textfont\msbfam=\tenmsb

\def\N{{\cal{N}}}
\def\R{{\fam\msbfam{R}}}

\lref\sundborg{
B.~Sundborg,
``The Hagedorn transition, deconfinement and $\N = 4$ SYM theory,''
Nucl.\ Phys.\ B {\bf 573}, 349 (2000)
[arXiv:hep-th/9908001].
}

\lref\DasguptaHX{
K.~Dasgupta, M.~M.~Sheikh-Jabbari and M.~Van Raamsdonk,
``Matrix perturbation theory for M-theory on a PP-wave,''
JHEP {\bf 0205}, 056 (2002)
[arXiv:hep-th/0205185].
}

\lref\DasguptaRU{
K.~Dasgupta, M.~M.~Sheikh-Jabbari and M.~Van Raamsdonk,
``Protected multiplets of M-theory on a plane wave,''
JHEP {\bf 0209}, 021 (2002)
[arXiv:hep-th/0207050].
}

\lref\MaldacenaRB{
J.~Maldacena, M.~M.~Sheikh-Jabbari and M.~Van Raamsdonk,
``Transverse fivebranes in matrix theory,''
JHEP {\bf 0301}, 038 (2003)
[arXiv:hep-th/0211139].
}

\lref\KimRZ{
N.~Kim, T.~Klose and J.~Plefka,
``Plane-wave matrix theory from $\N = 4$ super Yang-Mills on $\R \times
S^3$,''
Nucl.\ Phys.\ B {\bf 671}, 359 (2003)
[arXiv:hep-th/0306054].
}

\lref\BeisertJJ{
N.~Beisert,
``The complete one-loop dilatation operator of $\N = 4$ super Yang-Mills
theory,''
Nucl.\ Phys.\ B {\bf 676}, 3 (2004)
[arXiv:hep-th/0307015].
}

\lref\ammpv{
O.~Aharony, J.~Marsano, S.~Minwalla, K.~Papadodimas and M.~Van Raamsdonk,
``The Hagedorn/deconfinement phase transition in weakly coupled large $N$
gauge theories,''
arXiv:hep-th/0310285.
}

\lref\sv{
M.~Spradlin and A.~Volovich,
``The one-loop partition function of $\N = 4$
SYM on $\R \times S^3$,''
arXiv:hep-th/0408178.
}

\lref\klose{
T.~Klose and J.~Plefka,
``On the integrability of large $N$ plane-wave matrix theory,''
Nucl.\ Phys.\ B {\bf 679}, 127 (2004)
[arXiv:hep-th/0310232].
}

\lref\kp{
N.~Kim and J.~Plefka,
``On the spectrum of pp-wave matrix theory,''
Nucl.\ Phys.\ B {\bf 643}, 31 (2002)
[arXiv:hep-th/0207034].
}

\lref\bmn{
D.~Berenstein, J.~M.~Maldacena and H.~Nastase,
``Strings in flat space and pp waves from ${\cal N} = 4$ super Yang Mills,''
JHEP {\bf 0204}, 013 (2002)
[arXiv:hep-th/0202021].
}

\lref\KimZG{
N.~Kim and J.~H.~Park,
``Superalgebra for M-theory on a pp-wave,''
Phys.\ Rev.\ D {\bf 66}, 106007 (2002)
[arXiv:hep-th/0207061].
}

\lref\fss{
K.~Furuuchi, E.~Schreiber and G.~W.~Semenoff,
``Five-brane thermodynamics from the matrix model,''
arXiv:hep-th/0310286.
}

\Title
{\vbox{
 \baselineskip12pt
\hbox{hep-th/0409178}
\hbox{NSF-KITP-04-109}
}}
{\vbox{
\centerline{Two-Loop Partition Function }
\vskip .5cm
\centerline{in the Planar Plane-Wave Matrix Model}
}}

\centerline{
Marcus Spradlin${}^\dagger$,
Mark Van Raamsdonk${}^\ddagger$ and
Anastasia Volovich${}^\dagger$}

\bigskip
\bigskip

\centerline{${}^\dagger$Kavli Institute for Theoretical Physics,
University of California}
\centerline{Santa Barbara, CA 93106 USA}
\centerline{\tt spradlin, nastja@kitp.ucsb.edu}

\smallskip

\centerline{${}^\ddagger$Department of Physics and Astronomy, University
of British Columbia}
\centerline{6224 Agricultural Road,
Vancouver, B.C., V6T 1W9, Canada}
\centerline{\tt mav@physics.ubc.ca}

\bigskip
\bigskip

\centerline{\bf Abstract}

We perform two independent calculations of the two-loop partition function for
the large $N$ 't Hooft limit of the plane-wave matrix model, conjectured to be
dual to the decoupled little string theory of a single spherical type IIA
NS5-brane.  The first is via a direct two-loop path-integral calculation in the
matrix model, while the second employs the one-loop dilatation operator of
four-dimensional $\N = 4$ Yang-Mills theory truncated to the $SU(2|4)$
subsector.  We find precise agreement between the results of the two
calculations.  Various polynomials appearing in the result have rather special
properties, possibly related to the large symmetry algebra of the theory or to
integrability. 

\bigskip

\Date{}

\newsec{Introduction}

A recent and fascinating addition to the cast of maximally supersymmetric
Yang-Mills theories has been the plane-wave matrix model \bmn.
This theory, a massive deformation of the BFSS Matrix Theory preserving all
32 supercharges, has been conjectured to describe (in a particular large
$N$ limit) M-theory on the maximally supersymmetric plane-wave solution of
eleven-dimensional supergravity. The theory turns out to be much more
tractable than the usual BFSS matrix model \DasguptaHX, allowing perturbative
computations at fixed
$N$ in the limit of large mass $m$ and harboring a
powerful symmetry algebra \refs{\kp, \DasguptaRU, \KimZG} that allows
extrapolation of some perturbative results into the strongly-coupled regime.
Among this reliable information at strong coupling is direct evidence that
certain vacuum states of the model describe spherical BPS transverse
M5-branes of M-theory \MaldacenaRB.

Recently \refs{\KimRZ,\klose}, there has emerged an interesting
connection between the plane-wave matrix model and four-dimensional
${\cal N}=4$ Super-Yang-Mills (SYM) theory on $S^3$. At the
classical level, the plane-wave matrix model
emerges directly from the ${\cal N}=4$ theory through a consistent
truncation \KimRZ\ that keeps only the modes invariant under a certain $SU(2)$
subgroup of the $SU(2,2|4)$ superconformal algebra (the ``$SU(2|4)$
subsector''). In the strict large $N$ 't Hooft
limit, the relationship
between these two theories extends further.
The set of states of the ${\cal N}=4$ theory built from these modes
forms a subsector which is closed under renormalization at one loop.
Surprisingly,
all one-loop corrections to the energies of these states in the
${\cal N} = 4$ theory match precisely with the one-loop correction
to the energies computed in the plane-wave matrix model after the
correct identification of couplings between the two theories
\refs{\KimRZ,\BeisertJJ}. The common one-loop Hamiltonian governing
these energy shifts corresponds to an integrable $SU(2|4)$ spin chain,
so at least at the one-loop level, the recently discovered integrability
properties of the planar ${\cal N} = 4$ SYM theory extend to the
plane-wave matrix model. In fact, recent explicit calculations
\klose\ in the plane-wave matrix model (for a certain closed subsector
of scalar modes) suggest that both the integrability properties and the
equivalence with the appropriate subsector (the ``$SU(2)$ subsector'')
of the ${\cal N}=4$ theory persist even to
three loops!

Given these results, it is natural to wonder whether the full plane-wave
matrix model is integrable in the 't Hooft large $N$ limit.
It is important to note that this limit is quite different from the Matrix
Theory limit conjectured to define M-theory on the plane wave (which
clearly has no chance of being integrable). In fact, it was argued in
\MaldacenaRB\ that the 't Hooft large $N$ limit, defined about the trivial
vacuum state of the theory, is a decoupling limit which only keeps the
excitations of a single spherical fivebrane. More precisely, since
this limit does not decompactify the M-theory circle, the fivebrane
should be interpreted as a spherical type IIA NS5-brane with the 't Hooft
parameter related to the sphere radius in units of $\alpha'$.
The fact that this limit describes the decoupled physics of only a single
brane (according to the arguments of \MaldacenaRB) provides
additional hope that it may indeed be integrable.

If the 't Hooft large $N$ limit of the plane-wave matrix model does turn out
to be integrable, one might aspire to calculate the exact spectrum
for all values of the coupling, or equivalently, to find an
analytic expression for the exact partition function as a function of
coupling. Motivated by the hope that such an expression exists, we
proceed in this note to calculate directly the leading terms in its
weak coupling expansion.
Thus, we compute the two-loop partition function
for the strict large $N$ limit of the plane-wave matrix model about
its trivial vacuum,\foot{At $N=\infty$ with fixed finite 't Hooft coupling, the free energy diverges at a finite temperature, and our results for the partition function are valid below this temperature. At large but finite N, the story is more complicated, since the model has of order $e^{\sqrt{N}}$ vacua which should all contribute since their ground state energies are all zero. Thus, our result should only be interpreted in the context of the strict large $N$ 't Hooft limit for which the vacua decouple.} extending the previously calculated \refs{\ammpv,\fss} 
zero-coupling result given by (2.6) below. We find that the correction to the partition function takes the form
\eqn\mainresult{
\delta\ln \Tr[e^{-\beta H}]
= {3 \lambda \over 4 \pi^2}
 \ln(y) \left\{ \sum_{n=1}^\infty {n g((-1)^{n+1} y^n) \over
1 - z((-1)^{n+1} y^n)} - {p(y) \over q(y)} \right\},
}
where $\lambda = g_{\rm YM}^2 N$ is the 't Hooft parameter of
the ${\cal N} = 4$ gauge theory (we relate it to the matrix model
parameters below),
$y=e^{-\beta m/12}$,
and $z$, $g$, $p$ and $q$ are polynomials in $y$ given below
in (2.5), (2.20) and (2.24). These have some rather special properties that we comment on in section 4.      

Our calculation is carried out by two independent methods. The first
method, in section 2, is a direct two-loop path-integral calculation
using the Euclidean matrix model action with Euclidean time
compactified on a circle of radius $\beta = 1/T$. Our second method,
described in section 3, amounts to an explicit sum over states of the
Boltzmann factor, taking into account the leading order energies together
with their one-loop corrections. For this approach, we use the one-loop
equivalence to the $SU(2|4)$ subsector of the ${\cal N}=4$ SYM theory,
and apply the general analysis of \sv\ to express the subleading
terms in the partition function in terms of the one-loop dilatation
operator of four-dimensional $\N = 4$ Yang-Mills theory truncated
to the $SU(2|4)$ subsector. While the details of the two calculations
look rather different, both calculations precisely give \mainresult.

Even in the limit of zero coupling (studied previously in
\refs{\fss, \ammpv}), this partition function displays interesting
Hagedorn behavior, with a limiting temperature in the strict large $N$
limit at which the free energy diverges logarithmically.\foot{For large
but finite $N$, this divergence signals a phase transition to a
deconfined phase with free energy of order $N^2$
\refs{\sundborg, \fss, \ammpv}.} This Hagedorn behavior is presumably
associated with the Little Strings of the decoupled IIA NS5-brane
defined by this limit. From our two-loop results, we can determine
the change in the Hagedorn temperature as the coupling is turned on
(i.e. as the sphere on which the Little Strings live grows from zero size),
and we find that it increases with the coupling for small $\lambda$.
This is consistent with the suggestion \MaldacenaRB\ that the strong
coupling limit should be equivalent to the free conformal theory
associated with a single spherical M5-brane, for which we expect
no Hagedorn behavior.

\newsec{Plane-Wave Matrix Model Path Integral}

In this section, we compute the two-loop partition function for the 't Hooft
limit of the plane-wave matrix model directly via a path-integral calculation.
We follow all matrix model conventions of \DasguptaHX, in which the
plane-wave matrix model action in Euclidean signature is given
by\foot{Scalar indices $i,j,k$ and $a,b,c$ are associated with the vector representations of $SO(3)$ and $SO(6)$ respectively, while fermion indices of $\alpha, \beta$ and $I,J$ are in the spinor representations of these groups. The $g$'s are Clebsch-Gordon coefficients relating the vector of $SO(6)$ to the antisymmetric product of two spinors (fundamentals of $SU(4)$). We set $l_{\rm P} = 1$, but we can restore $l_{\rm P}$ in any
formulae using the fact 
that $R$ and $1/m$ have dimensions of length.} 
\eqn\action{\eqalign{
{\cal L} &=  \Tr \left( {1 \over 2} (D_0 {X}^i)^2 + {m^2 \over 18} (X^i)^2 
+{1 \over 2} (D_0 X^a)^2 + {m^2 \over 72}  (X^a)^2 + i \psi^{\dagger I \alpha} D_0 \psi_{I \alpha} + {m \over 4} 
\psi^{\dagger I \alpha} \psi_{I \alpha}\right) \cr 
& \qquad + \; R^{3 \over 2} \Tr \left( {i m  \over 3} \epsilon_{ijk} X^i 
X^j X^k + \psi^{\dagger I \alpha} \sigma^i_\alpha {}^\beta [X^i, \psi_{I \beta}]
\right. \cr
& \left. \qquad \qquad \qquad \qquad  
- {1 \over 2} \epsilon_{\alpha \beta} \psi^{\dagger \alpha I} g^a_{IJ} 
[X^a, \psi^{\dagger \beta J}] + {1 \over 2} \epsilon^{\alpha \beta} 
\psi_{\alpha I} g^{ \dagger a IJ} [X^a, \psi_{\alpha J}] \right) \cr
& \qquad - \; R^3 \Tr \left(  { 1 \over 4} [X^i , X^j]^2 
+ {1 \over 4} [X^a, X^b]^2 + { 1 \over 2} [ X^i, X^a]^2 \right).
}}

Choosing the gauge $\partial_t A_0 = 0$, and introducing the corresponding
Fadeev-Popov determinant $\Delta$, the thermal partition function is
\eqn\partdef{
Z = \int [dX^i] [dX^a][d \psi_{I\alpha}] [dA_0] \Delta
e^{-\int_0^\beta dt\,{\cal L}_{\rm Euc}}
}
where the time direction has been compactified with radius $\beta = 1/T$ and
bosons/fermions are taken to have periodic/antiperiodic boundary conditions
respectively.  As explained in \refs{\fss,\ammpv}, all modes are effectively
very massive at weak coupling except the zero mode of the gauge field on the
thermal circle.  It is then convenient as an intermediate step to integrate out
all other modes to produce an effective action for this zero-mode, which we
denote by $\alpha$.  As argued in \refs{\ammpv}, the resulting effective action
for this mode may depend only on the unitary matrix $U = e^{i \beta \alpha}$
(the Wilson line of the gauge field around the thermal circle), and the effect
of the determinant $\Delta$ in the path integral is precisely to convert the
measure $[dA_0]$ into the Haar measure $[dU]$ for unitary matrices.  Thus, the
partition function reduces to an ordinary unitary one-matrix model.

\subsec{One-Loop Result}

The evaluation of the partition function to one 1-loop has been carried out in
\refs{\fss,\ammpv}.  The result is
\eqn\oneloop{
Z_{\rm{1-loop}} = \int [dU] e^{-S^{\rm eff}_{\rm 1-loop}(U)}
}
where
\eqn\aaa{
S^{\rm eff}_{\rm 1-loop}(U) = \sum_{n=1}^\infty {1 \over n} z((-1)^{n+1} y^n)
\Tr(U^n) \Tr(U^{\dagger n}).
}
Here, we define $y = e^{- \beta m / 12}$ and
\eqn\zdef{
z(y) = 6 y^2 + 8 y^3 + 3 y^4
}
is the single mode (letter) partition function.  At strictly infinite $N$, the
free energy has a Hagedorn divergence at $T_{\rm H} = m/(12 \ln 3)$.  Below
this temperature, the model is governed by a stable saddle point for which the
eigenvalues of $U$ are spread uniformly around the unit circle, so that
$\Tr(U^n) = 0$.  Performing the Gaussian integral about this configuration
gives the first non-zero contribution to the free energy, and we obtain
\eqn\explicit{
Z_{\rm 1-loop} = e^{-\beta F_{\rm 1-loop}} = \prod_{n=1}^\infty
{1 \over 1 - z((-1)^{n+1} y^n)}.
}

\subsec{Two-Loop Calculation}

At two loops, the partition function is given by
\eqn\twoloop{
Z_{\rm 2-loops} = \int [dU]
e^{-S^{\rm eff}_{\rm 1-loop}(U)-S^{\rm eff}_{\rm 2-loop}(U)},
}
where
\eqn\sefftwo{\eqalign{
e^{-S^{\rm eff}_{\rm 2-loop}} &= \langle e^{-S_{\rm int}} \rangle_{\rm 2-loop}
\cr
& = \exp\left( -\langle S_4 \rangle +\half  \langle S_3 S_3 \rangle\right),
}}
where $S_3$ and $S_4$ are the cubic and quartic terms in the
action \action.
Here, the expectation values are evaluated in the free theory with fixed
 $\alpha$.

The required propagators follow from the quadratic action
\eqn\quadr{
S_2 = \int d t \; \Tr \left( {1 \over 2 } X^i(-D_0^2  + {m^2 \over 9})
X^i + {1 \over 2 } X^a(-D_0^2  + {m^2 \over 36}) X^a +
\psi^{\dagger I \alpha}( D_0 + {m \over 4}) \psi_{I \alpha} \right).
}
For the boson propagators, we find 
\eqn\props{\eqalign{
\langle X^i_{pq} (t) X^j_{rs} (t') \rangle &= \delta^{ij}
\Delta_{m \over 3}(t-t',\alpha)^{ps,rq},\cr
\langle X^a_{pq} (t) X^b_{rs} (t') \rangle &= \delta^{ab}
\Delta_{m \over 6}(t-t',\alpha)^{ps,rq}.
}}
The propagator $\Delta$ is defined to be a periodic function of $t$ given in
the domain $[0,\beta)$ by
\eqn\bprop{
\Delta_M(t,\alpha) = {e^{i \alpha t} \over 2 M} \left(
{e^{-M t} \over 1 - e^{i \alpha \beta} e^{-M \beta}}
- {e^{Mt} \over 1 - e^{i \alpha \beta} e^{M \beta}} \right),
}
where $\alpha$ is short for $(\alpha \otimes 1) - (1 \otimes \alpha)$ and
matrix indices have been suppressed.  The fermion propagator is
\eqn\fprops{
\langle (\psi_{I \alpha}(t))_{pq} (\psi^{\dagger J \beta})(t')\rangle
= \delta^J_I \delta^\beta_\alpha \Delta^{\rm F}_{m \over 4}(t-t',\alpha)^{ps,rq},
}
where $\Delta^{\rm F}$ is defined to be an antiperiodic function of $t$ given
in the domain $[0, \beta)$ by
\eqn\fprop{
\Delta^F_M(t,\alpha) =
e^{i \alpha t} {e^{-M t} \over 1 - e^{i \alpha \beta} e^{-M \beta}}.
}

There are six correlators contributing to $S^{\rm eff}_{\rm 2-loop}$. These are
\eqn\corrsone{
s_1 = \langle  - {R^3 \over 4} \int dt \; \Tr([X^i,X^j]^2) \rangle,
}
\eqn\aaa{
s_2 = \langle  - {R^3 \over 2} \int dt \; \Tr([X^i,X^a]^2) \rangle,
}
\eqn\aaa{\eqalign{
s_3 &= \langle  - {R^3 \over 4} \int dt \; \Tr([X^a,X^b]^2) \rangle \cr
s_4 &= \langle {R^3 m^2 \over 18} \int dt \; \Tr(\epsilon^{ijk} X^i X^j X^k)
\int d t'\; \Tr(\epsilon^{lmn} X^l X^m X^n) \rangle \cr
s_5 &= \langle -{R^3 \over 2} \int dt \; \Tr( \psi^{\dagger I \alpha}
\sigma^i_\alpha {}^\beta [X^i, \psi_{I \beta}]) \int
dt' \; \Tr( \psi^{\dagger I' \alpha'} \sigma^{i'}_{\alpha'} {}^{\beta'}
[X^{i'}, \psi_{I' \beta'}]) \rangle \cr
s_6 &= \langle {R^3 \over 4} \int dt \; \Tr(\epsilon_{\alpha \beta}
\psi^{\dagger I \alpha} g_{IJ}^a [X^a, \psi^{\dagger J \beta}])
\int dt' \; \Tr(\epsilon^{\alpha' \beta'} \psi_{I' \alpha'}
g^{\dagger a' I'J'} [X^{a'}, \psi_{J' \beta'}]) \rangle.
}}

These correlators contribute to the partition function in two different ways.
First, planar diagrams contribute terms to the double-trace effective action
for $U$ which modify the Gaussian integral and result in order $\lambda$
corrections to the denominators in \explicit.  Since the Gaussian integral is
actually the subleading contribution to the large $N$ free energy (the leading
${\cal O}(N^2)$ contribution vanishes for this saddle point) there are also
contributions at the same order arising from {\it nonplanar} two-loop diagrams.\foot{We thank Ofer Aharony for emphasising that non-planar diagrams must play a role here.}
These are independent of $U$ and give a temperature-dependent prefactor to the
infinite product in \explicit. 

\subsec{Planar Contribution}

{}From the six correlators above, we first write the planar contributions,
giving in the first line the complete expression for the planar part of the
correlator in terms of propagators and in the second line, the terms
contributing to the double trace action. There are in addition three-trace
terms, but these do not contribute to the partition function at infinite $N$. We find:
\eqn\planarone{\eqalign{
(s_1)_{\rm pl} &= 3 \beta R^3 \Delta_{m \over 3}(0, \alpha_{ab}) \Delta_{m \over 3}(0, \alpha_{ac}) \cr
&\rightarrow  \sum_{n=1}^\infty {27 \over 2} \beta {R^3 N \over m^2} (y^{8n}+2 y^{4n}) \Tr(U^n) \Tr(U^{\dagger n}) \cr
(s_2)_{\rm pl} &= 18 \beta R^3 \Delta_{m \over 3}(0, \alpha_{ab}) \Delta_{m \over 6}(0, \alpha_{ac}) \cr
&\rightarrow  \sum_{n=1}^\infty 162  \beta {R^3 N \over m^2} (y^{6n} + y^{4n}+ y^{2n}) \Tr(U^n) \Tr(U^{\dagger n}) \cr
(s_3)_{\rm pl} &= 15 \beta R^3 \Delta_{m \over 6}(0, \alpha_{ab}) \Delta_{m \over 6}(0, \alpha_{ac}) \cr
&\rightarrow  \sum_{n=1}^\infty 270 \beta {R^3 N \over m^2} (y^{4n}+ 2 y^{2n} ) \Tr(U^n) \Tr(U^{\dagger n})
}}
\eqn\aaa{\eqalign{ (s_4)_{\rm pl} &=  -m^2 \beta R^3 \int dt \Delta_{m \over 3}(t, \alpha_{ab}) \Delta_{m \over 3}(t, \alpha_{bc}) \Delta_{m \over 3}(t, \alpha_{ca}) \cr
&\rightarrow  \sum_{n=1}^\infty {81 \over 2} \beta {R^3 N \over m^2} (y^{8n} - 2 y^{4n} ) \Tr(U^n) \Tr(U^{\dagger n}) \cr
(s_5)_{\rm pl} &=  -24 \beta R^3 \int dt \Delta_{m \over 3}(t, \alpha_{ab}) \Delta^{\rm F}_{m \over 4}(t, \alpha_{bc}) \Delta^{\rm F}_{m \over 4}(\beta - t, \alpha_{ac}) \cr
&\rightarrow  \sum_{n=1}^\infty 216 \beta {R^3 N \over m^2} (-1)^{n+1} (y^{6n} + 2 y^{5n} + 2 y^{3n} - 3 y^{2n}) \Tr(U^n) \Tr(U^{\dagger n}) \cr
(s_6)_{\rm pl} &=  -48 \beta R^3 \int dt \Delta_{m \over 6}(t, \alpha_{ab}) \Delta^{\rm F}_{m \over 4}(t, \alpha_{bc}) \Delta^{\rm F}_{m \over 4}(t, \alpha_{ca}) \cr
&\rightarrow \sum_{n=1}^\infty 216 \beta {R^3 N \over m^2} (-1)^{n+1} (y^{7n} - y^{3n}) \Tr(U^n) \Tr(U^{\dagger n}).
}}
In the expressions above, each of the propagators contribute factors of $\alpha$ to two of the three index loops, which we label by $a$,$b$, and $c$. The notation $\alpha_{ab}$ indicates that for the tensor products $(\alpha \otimes 1) - (1 \otimes \alpha)$ appearing in the propagator, the first and second elements of the tensor product appear in the traces associated with index loops $a$ and $b$ respectively.

Combining all terms, we find that the two-loop contribution to the double-trace
effective action for $U$ is
\eqn\plresult{
S^{\Tr^2}_{\rm 2-loop}(U) = - \widetilde{\lambda} \sum_{n=1}^\infty
\ln(y) g((-1)^{n+1} y^n) \Tr(U^n) \Tr(U^{\dagger n}),
} 
where 
\eqn\gydef{
g(y) = y^2(1+y)^4(1+y^2)
}
and we have defined a 't Hooft coupling $\widetilde{\lambda} = 648 R^3 N/m^3$.

\subsec{Nonplanar Contribution}

We now evaluate the nonplanar contributions from the six correlators above.
In this case, since there is only a single index loop and since each term in
the propagators contributes an equal number of $U$s and $U^\dagger$s, we will
always end up with just the identity matrix inside the single trace.  Thus, the
same result will be obtained by setting $\alpha=0$ ($U=1$) in all propagators
from the start.  The nonplanar contributions are thus
\eqn\nonplanarone{\eqalign{
(s_1)_{\rm np} &= -3 \beta R^3 \Delta_{m \over 3}(0, 0) \Delta_{m \over 3}(0, 0) \cr
&=   -{27 \over 4} \beta {R^3 N \over m^2} {(1+y^4)^2 \over (1-y^4)^2},
}}
\eqn\aaa{\eqalign{
(s_2)_{\rm np} &=  -18 \beta R^3 \Delta_{m \over 3}(0, 0) \Delta_{m \over 6}(0, 0) \cr
&=  -81 \beta {R^3 N \over m^2} {(1+y^4)(1+y^2)\over (1-y^4)(1-y^2)} \cr
(s_3)_{\rm np} &=  -15 \beta R^3 \Delta_{m \over 6}(0, 0) \Delta_{m \over 6}(0, 0) \cr
= & -135 \beta {R^3 N \over m^2} {(1+y^2)^2 \over (1-y^2)^2} \cr
(s_4)_{\rm np}
&=  m^2 \beta R^3 \int dt \Delta_{m \over 3}(t, 0) \Delta_{m \over 3}(t, 0) \Delta_{m \over 3}(t, 0) \cr
&= {27 \over 4} \beta {R^3 N \over m^2} {(1+10y^4+y^8) \over (1-y^4)^2} \cr
(s_5)_{\rm np} &=  24 \beta R^3 \int dt \Delta_{m \over 3}(t, 0) \Delta^{\rm F}_{m \over 4}(t, 0) \Delta^{\rm F}_{m \over 4}(\beta - t, 0) \cr
&=  216 \beta {R^3 N \over m^2} {y^3 \over (1+y^3)^2} \cr
(s_6)_{\rm np} &=  48 \beta R^3 \int dt \Delta_{m \over 6}(t, 0) \Delta^{\rm F}_{m \over 4}(t, 0) \Delta^{\rm F}_{m \over 4}(t, 0) \cr
&=  216 \beta {R^3 N \over m^2} {(1+2y^2-2y^6-y^8) \over (1+y^3)^2 (1-y^2)}.
}}
Combining all contributions, we find
\eqn\npresult{
S_{\rm np} = \widetilde{\lambda} \ln(y) {p(y) \over q(y)}
}
where $\widetilde{\lambda}$ is defined as above and
\eqn\pqdef{\eqalign{
p(y) &= y^2(1+y+y^2)(1+y+6y^2+y^3+6y^4+y^5+y^6),\cr
q(y) &= (1-y)^2(1+y)^2(1-y+y^2)^2 (1+y^2)^2.
}}

\subsec{Summary:  Two-Loop Result}

Using the results above, it is now straightforward to complete the calculation
of the two-loop partition function by performing the Gaussian integral around
the $\Tr(U^n) = 0$ saddle point.  Combining the one- and two-loop effective
actions for $U$, we find that the terms quadratic in traces are
\eqn\aaa{
S^{\Tr^2}_{\rm eff}(U) = 
\sum_{n=1}^\infty {1 \over n} \left(1 - z((-1)^{n+1} y^n) -
\widetilde{\lambda} \ln(y^n) g((-1)^{n+1} y^n) +
{\cal O}(\widetilde{\lambda}^2)\right) \Tr(U^n) \Tr(U^{\dagger n}).
}
{}From this, we can read off the appropriate modification of the denominators
in \explicit, so combining the results of the Gaussian integral with the
prefactor coming from the nonplanar diagrams, we obtain our final result
\eqn\partfn{
Z = e^{- \widetilde{\lambda} \ln(y) p(y) / q(y)} \prod_{n=1}^\infty {1 \over 1 - z((-1)^{n+1} y^n) - \widetilde{\lambda} \ln(y^n) g((-1)^{n+1} y^n)} + {\cal O}(\widetilde{\lambda}^2),
}
where $g$, $p$, and $q$ were defined in \gydef\ and \pqdef .
For what follows it will be convenient to write the first
${\cal{O}}(\widetilde{\lambda})$ correction
as
\eqn\result{
\delta \ln Z
= \widetilde{\lambda} \ln(y) \left\{ \sum_{n=1}^\infty {n g((-1)^{n+1} y^n) \over 1 - z((-1)^{n+1} y^n)} - {p(y) \over q(y)} \right\}.
}

\newsec{Dilatation Operator in the $SU(2|4)$ Subsector}

In this section we obtain the partition function \result\ from
a one-loop calculation in $\N = 4$ SYM gauge theory on $\R \times S^3$
by making use of the one-loop
isomorphism between the plane-wave matrix model and the $SU(2|4)$ subsector of
the gauge theory.  This subsector consists of those operators
built out of the six scalar fields $\phi$, the eight positive
chirality spinors $\lambda$, and the three self-dual components of the
field strength tensor $F$.
Henceforth we will use the matrix model notation in referring to these
modes respectively as $X^a$, $\psi_{I \alpha}$, and $X^i$.
Covariant derivatives of these fields do not
appear in this subsector.

The free Hamiltonian $H_0$ of the matrix model is
identified with the tree-level dilatation operator $D_0$ in
this subsector of gauge theory
according to
\eqn\aaa{
H_0 = {m \over 6} D_0.
}
Quantum corrections to the matrix model effective Hamiltonian
can be
computed for large $m$ via ordinary degenerate
quantum mechanical perturbation theory in the parameter $1/m^3$.
The first order correction has been shown \refs{\KimRZ,\BeisertJJ}
to agree with
the one-loop correction to the gauge theory dilatation operator,
\eqn\hamiltonians{
H_0 + {1 \over m^3} V_1^{\rm eff}
 = {m \over 6} \left(D_0 + {\lambda \over 4 \pi^2} D_2\right),
}
where $\lambda = g_{\rm YM}^2 N$ is the gauge theory `t Hooft parameter.
Although the $SU(2|4)$ subsector of the gauge theory is not closed under
renormalization beyond one loop, we can
take $H = (m/6) D$ in calculating the partition function
\eqn\aaa{
Z = \Tr[e^{-\beta H}] = \Tr[y^{2 D}] + {\cal{O}}(\lambda^2)
}
since at the moment
we are not interested in the higher order terms.

To calculate the leading term $\Tr[y^{2 D_0}]$ in
the partition function one simply has to
enumerate the operators appearing in the $SU(2|4)$ subsector
weighted by their bare dimension.  A straightforward
application of P\'olya theory yields the result \explicit.
The first correction to $\Tr[y^{2 D_0}]$ was studied in \sv,
where a combinatorial analysis of
the anomalous dimensions of gauge theory operators revealed that the
result has the general structure
\eqn\polya{
\delta \ln Z = {\lambda \ln(y) \over 2 \pi^2} 
\sum_{n=1}^\infty\left[{n \langle D_2((-1)^{n+1} y^n)\rangle
\over 1 - z((-1)^{n+1} y^n)} + \sum_{m=1}^\infty
\langle PD_2((-1)^{m+1} y^m,(-1)^{n+1}y^n)\rangle
\right].
}
This formula is valid for temperatures below the Hagedorn temperature at which the $N=\infty$ free energy diverges, as evidenced by the appearence of a pole at $y = 1/3$ in
the first term of \polya. Intriguingly, it is already apparent that this expression has
strong similarities with \result.

The quantities $\langle D_2\rangle$ and $\langle P D_2 \rangle$
are defined as follows.  The one-loop dilatation operator $D_2$ only
acts on two neighboring fields in any single-trace operator,
\eqn\aaa{
D_2 \Tr[A_1 \cdots A_L] = \sum_{i=1}^L \Tr[ A_1 \cdots D_2 (A_i 
A_{i+1}) \cdots A_L].
}
In the $SU(2|4)$ subsector, each letter $A_i$ corresponds to one of
the 17 fields $\{X^a, \psi_{I \alpha}, X^i\}$, so we can think
of $D_2$ as a $289 \times 289$ matrix acting on a pair of
letters $|A_1 A_2\rangle$.
The ingredients appearing in \polya\ 
are just traces of this matrix,
\eqn\traces{
\langle D_2(y) \rangle = \Tr[y^{2 D_{0(1)} + 2 D_{0(2)}} D_2], \qquad
\langle P D_2(u,w) \rangle = \Tr[u^{2 D_{0(1)}} w^{2 D_{0(2)}} P D_2],
}
where $P|A_1 A_2\rangle = (-1)^{F_1 F_2}
|A_2 A_1\rangle$ is the (graded) permutation operator
and $D_{0(i)}$ gives the bare dimension of the $i$th letter.

The one-loop dilatation operator $D_2$ in the $SU(2|4)$ subsector
takes the form
\eqn\dtwo{\eqalign{
D_2|X^a X^b\rangle&= {1 \over 2} |X^a X^b\rangle
- {1 \over 2} |X^b X^a\rangle + {1 \over 4} \delta^{ab} \delta^{cd}
 |X^c X^d\rangle,\cr
D_2|\psi_{I \alpha} X^i\rangle &=
|\psi_{I \alpha} X^i \rangle + {i \over 4} \epsilon^{ijk}
\sigma^j{}_\alpha{}^\beta \Big( |\psi_{I \beta} X^k\rangle
+ |X^k \psi_{I \beta} \rangle\Big),
\cr
D_2|X^i \psi_{I \alpha}\rangle &=
|X^i \psi_{I \alpha} \rangle + {i \over 4} \epsilon^{ijk}
\sigma^j{}_\alpha{}^\beta \Big( |\psi_{I \beta} X^k\rangle
+ |X^k \psi_{I \beta} \rangle\Big),\cr
D_2|X^a X^i\rangle &= {3 \over 4} |X^a X^i\rangle
- {1 \over 4} |X^i X^a\rangle +
{1 \over 2} g^{\dagger a IJ} \sigma^{i\alpha\beta}
\Big(
|\psi_{I\alpha} \psi_{J\beta}\rangle
- |\psi_{J \beta} \psi_{I \alpha}\rangle
\Big),}}
\eqn\aaa{\eqalign{
D_2|X^i X^a\rangle &= {3 \over 4} |X^i X^a\rangle
- {1 \over 4} |X^a X^i\rangle +
{1 \over 2} \sigma^{i\alpha\beta}
g^{\dagger a IJ} \Big(
|\psi_{I\alpha} \psi_{J \beta}\rangle
- |\psi_{J \beta} \psi_{I \alpha}\rangle\Big),\cr
D_2|\psi_{I \alpha} \psi_{J \beta}\rangle &=
{3 \over 4} |\psi_{I \alpha} \psi_{J \beta}\rangle
+ {1 \over 4} |\psi_{J \beta} \psi_{I \alpha}\rangle +
{1 \over 4} |\psi_{I \beta} \psi_{J \alpha}\rangle -
{1 \over 4} |\psi_{J \alpha} \psi_{I \beta}\rangle\cr
&\qquad\qquad- {1 \over 32} g^a_{IJ}
\sigma^i_{\alpha \beta}\Big(|X^aX^i\rangle
+ |X^iX^a\rangle\Big),\cr
D_2|\psi_{I \alpha} X^a \rangle &=
{5 \over 8} |\psi_{I \alpha} X^a \rangle
- {3 \over 8} |X^a \psi_{I \alpha}\rangle
+ {1 \over 8} g^{ab}{}_{I}{}^{J}\Big(|\psi_{J\alpha} X^b\rangle
+ |X^b \psi_{J\alpha}\rangle \Big),\cr
D_2|X^a \psi_{I \alpha} \rangle &=
{5 \over 8} |X^a \psi_{I \alpha}\rangle
- {3 \over 8} |\psi_{I \alpha} X^a \rangle
+ {1 \over 8} g^{ab}{}_I{}^J\Big(|\psi_{J\alpha} X^b\rangle
+ |X^b \psi_{J\alpha}\rangle \Big),\cr
D_2|X^i X^j\rangle &= {5 \over 4} |X^i X^j\rangle + {1 \over 4}
|X^j X^i\rangle - {1 \over 2} \delta^{ij} \delta^{kl} |X^k X^l\rangle.
}}
We obtained these expressions by restricting the general result written in
\BeisertJJ\ to the $SU(2|4)$ subsector.
The first line, familiar as the Hamiltonian
of the $SO(6)$ spin chain,
has been reproduced by a direct
calculation in plane-wave matrix perturbation theory \KimRZ.
It would be interesting to extend their analysis to the
full $SU(2|4)$ subsector.
The last line is the standard spin-1 $SU(2)$ spin chain Hamiltonian.
The required traces \traces\ can be easily read off from these
formulas, and we find
\eqn\traceresults{
\eqalign{
\langle D_2(y) \rangle &=
{3 \over 2} y^4 (1 + y)^3 (11 + 7 y),\cr
\langle P D_2(u,w) \rangle &=
-{9 \over 2} u^2 w^2 (1+u)(1+w) (3 + u +w - u w).
}}
It remains to plug \traceresults\ into \polya.
The first trace can be broken into the two terms
\eqn\why{
\langle D_2(y) \rangle = {3 \over 2} g(y) - {3 \over 2} (1 - z(y)) y^2 (1 + 4
y + 2 y^2),
}
where $g(y)$ is the same function we defined in 
\gydef.  The factor of $1-z$ in the second term cancels the denominator
in \polya, allowing this term to be summed explicitly, giving
\eqn\termone{
\sum_{n=1}^\infty{n \langle D_2((-1)^{n+1} y^n)\rangle
\over 1 - z((-1)^{n+1} y^n)}
= {3 \over 2} \sum_{n=1}^\infty{n g((-1)^{n+1} y^n)
\over 1 - z((-1)^{n+1} y^n)}
+ T_1(y)
}
with
\eqn\aaa{
T_1(y) = - {3 y^2(1+y+y^2)(1+y-3 y^2+4 y^3 - 3 y^4 + y^5 + y^6)
\over 2 (1-y)^2 (1+y)^2 (1+y^2)^2 (1-y+y^2)^2}.
}
The double sum of the second trace
in \polya\ can be evaluated explicitly,
giving the contribution
\eqn\termtwo{
T_2(y)
= -{9 y^4 (1+y+y^2)(3-y+3 y^2)\over 2(1-y)^2 (1+y)^2 (1+y^2)^2 (1 - y+y^2)^2}.
}
Remarkably, $T_1(y) + T_2(y) = - p(y)/q(y)$ (defined
in \pqdef), so
combining all terms yields the final expression
\eqn\resulttwo{
\delta \ln Z = {3 \lambda \ln(y) \over 4 \pi^2}
\left\{\sum_{n=1}^\infty {n g((-1)^{n+1} y^n)
\over
 1 - z((-1)^{n+1} y^n)} - {p(y) \over q(y)} \right\},
}
in precise agreement with \result.
The overall coefficient agrees after we make use
of $\lambda = (4 \pi^2 /3) \widetilde{\lambda}$,
which follows from the familiar relation
\eqn\aaa{
\left( {m \over 3 R}\right)^3 = {32 \pi^2 \over g_{\rm YM}^2}
}
between the matrix model and Yang-Mills parameters.

\newsec{Discussion}

The result we have derived 
contains a significant amount of physical information about the model. Specifically, the coefficient of $y^{n} \ln(y)$ in \partfn\ gives $12/m$ times the sum of one-loop corrections to the energies of all states with energy $n m / 12$ at zero coupling.  
These energy shifts result in a shift in the Hagedorn temperature at which the partition function diverges, and using \result\ we find that the 
corrected Hagedorn temperature
\eqn\hagedorn{
T_{\rm H} = {m \over 12 \ln 3} \left(
1 + {10 \over 81 \pi^2}
\lambda + {\cal{O}}(\lambda^2)
\right),
}
increases as we move to stronger coupling.

A particularly mysterious feature of our analysis is that both methods of calculation split naturally 
into two parts, yet the pieces on the two sides are not in direct correspondence. The two quantities $g$ and $p/q$ have quite distinct
origins (from planar versus non-planar
diagrams) in the calculation of section 2,
while $\langle D_2 \rangle$ and $\langle P D_2 \rangle$ have
similarly distinct interpretations in section 3. However, the crucial equation \why\ shows that
there is not a direct identification between $g$ and $\langle D_2
\rangle$. It would be very interesting to find a direct interpretation for
$g$, or equivalently, to determine whether there is a sense
in which the decomposition \why\ is natural, from the spin chain viewpoint.

Finally, it would be interesting to understand whether any features of the results we have derived are related to the special properties of the model, such as its large superalgebra or integrability properties. The overall form of the expression \mainresult\ is rather generic for large $N$ gauged matrix models
in 0+1 dimensions, so any special features should show up in the polynomials $g$,$p$, and $q$ themselves. In fact, these polynomials are quite non-generic:
they possess
a significant degree of factorization, a symmetry
under reversing the order of exponents, and, in the case of $g$ and $q$,
all nonzero roots lie on the unit circle. Whether any of these features, some of which 
appear also in the full ${\cal N} = 4$ SYM result \sv, relate to integrability or supersymmetry is left as a question for future work.

\bigskip

\centerline{\bf Acknowledgments}\nobreak

We would like to thank O.~Aharony, S.~Minwalla, J.~Plefka, R.~Roiban, and G.~Semenoff for helpful discussions.
This research was supported in part by the National Science Foundation under
Grant No.~PHY99-07949 (MS, AV), and in part by the Natural Sciences and Engineering Council of Canada and by the Canada
Research Chairs programme (MVR).

\listrefs

\end